# Complex dielectric and impedance behavior of magnetoelectric $Fe_2TiO_5$

Shivani Sharma[1], Tathamay Basu[2], Aga Shahee[1], K. Singh[1*], N. P. Lalla[1] and E. V. Sampathkumaran[2]

[1]*UGC-DAE Consortium for Scientific Research, University Campus, Khandwa Road, Indore - 452001, India*

[2]*Tata Institute of Fundamental Research, Homi Bhabha Road, Colaba, Mumbai- 400005, India*

**ABSTRACT**

We have investigated the complex dielectric and impedance properties of magnetoelectric compound $Fe_2TiO_5$ (FTO) as a function of temperature ($T$) and frequency ($f$) to understand the grain ($G$) and grain boundary ($G_b$) contributions to its dielectric response. The temperature and frequency dependent dielectric permittivity ($\varepsilon'$) data shows a sharp increase in permittivity above 200K accompanied with a frequency dependent peak in $\tan\delta$. At $T$<175K, only $G$ contribution dominates even at lower frequency (~100Hz), but for $T \geq 175K$, the $G_b$ contribution starts appearing at low frequency. The value of critical frequency distinguishing these two contributions increases with increasing temperature. The observed non-Debye dielectric relaxation follows thermally activated process and is attributed to polaron hopping. Further the frequency dependence of ac conductivity follows the Jonscher's power-law. The temperature dependency of critical exponent '$s$' shows that the correlated barrier hopping model is appropriate to define the conductivity mechanism of FTO in the studied temperature regime.

# I. INTRODUCTION

The transition metal oxides have drawn great attention due to their various exotic properties such as colossal magnetoresistance, superconductivity, multiferroicity, multiglass and high dielectric constant ($\varepsilon'$) etc. The high dielectric constant materials are required for many technologically important applications because of their potential in energy storage devices[1,2]. For device applications, the dielectric materials should have low dissipation factor (loss) and small frequency ($f$) and temperature ($T$) dependence so that they can be utilized over broad frequency and temperature range. The observation of high dielectric constant in $CaCu_3Ti_4O_{12}$ (CCTO) enhanced the impetus to search new such high dielectric constant oxide materials[3]. As a result, many such studies[4-10] have been made to explore the dielectric behavior of different new systems. However, the high dielectric constant of different materials may have different origins such as internal barrier layer capacitance, Maxwell-Wagner effect, orbital excitations etc.[4,11,12].

Among transition metal oxides, some titanates exhibit ferroelectric/relaxor properties due to the off-centering of $Ti^{4+}$ ion[13-15]. Moreover, "*multiglass*" concept, first proposed for titanates[16] is subsequently observed in many other systems[9,17-19]. $Fe_2TiO_5$ (FTO) is a well-studied uniaxial anisotropic spin-glass insulator with two successive glassy freezing temperatures i.e. transverse ($T_{TF}$ = 9K) and longitudinal ($T_{LF}$ = 55 K)[20]. Apart from its interesting magnetic properties, recently, we have explored the magnetoelectric coupling and multiglass behavior of this system in the vicinity of magnetic ordering temperature[21]. However, its dielectric and transport behavior well above $T_{LF}$ has not been yet explored explicitly. There are only few reports on its dielectric studies[21-23] without any detail investigation on its conduction mechanism. Iwauchi et al.[22] have studied the dielectric properties to investigate the interfacial polarization due to heterogeneity of different combinations i.e. (FTO and $TiO_2$) and (FTO and $\alpha$-$Fe_2O_3$) etc. The ac electrical measurements such as dielectric permittivity, impedance spectroscopy and electric modulus are the important tools to investigate the detail electrical properties, e.g. the conduction mechanism like polaronic-hopping[24] of the materials. Such measurements are very advantageous to distinguish the grain ($G$) and grain boundary ($G_b$) contributions in polycrystalline materials, which helps in characterizing the materials for desired applications.

In the present study, we have explored the frequency (100Hz-1MHz) and temperature (150-300K) dependence of dielectric permittivity, ac impedance and electric modulus of FTO with the prime motivation to understand the $G$ and $G_b$ contributions and effective conduction mechanism in this frequency and temperature regime. Our comprehensive and quantitative analysis shows that FTO exhibits non-Debye type dielectric relaxation in the studied temperature range and at higher temperatures ($T>175K$), the transport mechanism is supported by polaronic hopping and explained by correlated barrier hopping (CBH) model.

## II. EXPERMENTAL

The polycrystalline sample was prepared via solid state reaction route and the details are already reported[21]. The same sample which was investigated in Ref. 21, has been used in the present study. The complex dielectric permittivity as a function of temperature was measured at different frequencies (1-100kHz) from 150K to 300K during warming (0.5K/min). The measurement was done using E4980A LCR meter (Agilent technologies) coupled with the Quantum Design physical property measurement system (PPMS). We have measured the real part of capacitance and loss (tan$\delta$) as a function of temperature at different frequencies and dielectric constant is derived from capacitance by using the thickness and area of the capacitor. We have tried our best to minimize the major source of error in the dielectric constant arises from the accurate dimensions measurements of the studied capacitor. Isothermal frequency dependent (till 1MHz) capacitance and tan$\delta$ was also measured at different temperatures, above and below $T_{max}$ (peak temperature in tan$\delta$) under different applied ac voltages (0.1-2V). Except 300K, we are showing the results measured at 1V ac amplitude. All other parameters (permittivity, impedance, and modulus and ac conductivity) are calculated as mentioned in the text of respective sections.

## III. RESULTS AND DISCUSSION

FTO crystallizes in (pseudobrookite) orthorhombic structure (*Cmcm*)[25-27]. The refined lattice parameters are found to be consistent with the earlier report[26]. The detailed–room temperature (RT) XRD results of this sample have already been given in our recent paper[21]. The structural arrangement of atoms in FTO is illustrated in Fig. 1. However, the best structural refinement of our XRD data has been done by considering partial disorder between Fe and Ti sites[21]. From Fig. 1, it can be noticed that the crystal structure of FTO contains two kinds of octahedra i.e. $TiO_6$ and $FeO_6$ and both these octahedra are distorted[28]. The distortion in $FeO_6$ octahedra is more as compared to $TiO_6$ octahedra. The $FeO_6$ octahedra share their edges with the first and second nearest $FeO_6$ octahedra and the third nearest $FeO_6$ octahedra is linked through corner sharing. These edge shared octahedra form tri-octahedra units which cause distortion in the crystal structure (see Fig. 1). However, the $TiO_6$ octahedra share its corner with other $TiO_6$ octahedron.

The temperature dependence of real and imaginary parts of the complex dielectric permittivity ($\varepsilon'$ and tan$\delta$) at different frequencies (5-100kHz) is presented in Fig. 2a and 2b, respectively. Figure 2 shows that up to 175K, there is no much frequency dispersion in $\varepsilon'$ and tan$\delta$. However, above 175K, $\varepsilon'$ increases sharply with increasing temperature accompanied by a peak in tan$\delta$. The peak in tan$\delta$ shifted towards higher temperature with increasing frequency which shows a clear dielectric relaxation behavior as observed for relaxor ferroelectric materials[29-31]. The peak height as well as its full width half maximum slightly increases with increasing frequency as observed in other systems also[6,13,19]. Near RT, the $\varepsilon'$ approaches ~$10^3$ and remains nearly frequency independent above 50kHz. To further explore the relaxation mechanism, the temperature of peak position at different frequencies ($\omega=2\pi f$) (from tan$\delta$ vs T) has been noted and relaxation time ($\tau$) has been calculated by using the formula, $\tau=1/\omega$. The error in deciding the temperature of peak position can be in mK. To calculate the activation energy ($E_a$), we have fitted the temperature dependence of $\tau$ with Arrhenius equation

$$\tau = \tau_o e^{[E_a/(K_B T)]} \qquad (1)$$

where $\tau_o$ is the pre-exponential factor, $K_B$ is the Boltzmann's constant (~$1.38\times10^{-23}$ $m^2.kg.s^{-2}.K^{-1}$) and $E_a$ is the activation energy[32]. The fitted curve is presented in the inset of Fig. 2b. The observed fitting parameters $\tau_o$ and $E_a$ are $\approx5(\pm0.6)\times10^{-11}$s and $\approx140(\pm2)$meV, respectively. The observed value of the activation energy is comparable with that reported for polaron hopping[32,33]. This sample is an insulator and having resistance in mega-ohm range at room temperature. This thermally activated process can be governed by polaron hopping. The sudden increase in $\varepsilon'$ at higher temperature ($T>200K$) is most probably due to enhancement of polaron hopping. In order to further understand the underlying relaxation behavior, we have performed isothermal frequency dependent measurements at different temperatures, see Fig. 3. Below 175K, the dispersion in $\varepsilon'$ vs $f$ is very small but becomes larger at higher temperature. At higher temperatures, $\varepsilon'$ first decreases with increasing frequency, resulting in an inflection point and tends to become almost frequency independent at higher frequencies (see Fig. 3a for 200K). At $T>200K$, after the inflection point, $\varepsilon'$ again decreases slowly at higher frequencies. The maxima in tan$\delta$ appeared at different frequencies with increasing temperature and this systematic shift can be seen clearly in the corresponding tan$\delta$ vs $f$ plots in Fig.3b.

In order to further investigate the origin of high dielectric constant (for T>200K), we have analyzed complex impedance ($Z^*=Z'+iZ''$) at different temperatures on both sides of the relaxation peak i.e. maxima in tan$\delta$. $Z^*$ is represented as $Z^* = -i\omega C_o \varepsilon^*$ and

$$Z' = \frac{1}{2\pi f C_o}\left[\frac{\varepsilon''}{\varepsilon'^2 + \varepsilon''^2}\right]; \quad Z'' = \frac{1}{2\pi f C_o}\left[\frac{\varepsilon'}{\varepsilon'^2 + \varepsilon''^2}\right] \quad (2)$$

where $\omega$ is the angular frequency (= $2\pi f$), $\varepsilon^*$, $\varepsilon'$, $\varepsilon''$ (= $\varepsilon'$tan$\delta$) are the complex, real and imaginary parts of the dielectric permittivity and $C_o$ is capacitance of vacuum[34]. The frequency dependence of real ($Z'$, resistive) and imaginary ($Z''$, reactive) part of complex impedance $Z^*$ at different temperatures is shown in Fig. 3 (c-d). The dielectric relaxation is also evident from Fig. 3d.

Further we are presenting the impedance results in the form of Nyquist plots, where one can distinguish between $G$ and $G_b$ contributions and even the electrode effect also (if present). In Fig. 4, the Nyquist plots are shown only at some selected temperatures to avoid the haziness in the figure. The presence of these contributions can be judged by seeing one or more arcs in a Nyquist plots[35]. The first and second arcs (from high frequency side) are attributed to $G$ and $G_b$ contributions, respectively, whereas the third arc (if present at lower frequency side) will be due to electrode effects[35,36]. As evident from $\varepsilon'$ vs $T$ results (Fig. 2a), at lower temperatures i.e. $T<160K$, the $\varepsilon'$ is almost frequency independent and dominated by $G$ contributions only. Hence to differentiate between $G$ and $G_b$ contribution, we are showing the Nyquist plots only for $T\geq160K$ in Fig. 4(a-f). At $T=160K$, we have observed a single semicircle (see Fig. 4a) which indicates that the measure frequency range the observed behavior is originating from $G$ contribution only. At $T\geq175K$, the second arc starts appearing approximately at 0.1kHz, see Fig. 4b. From Fig. 4(a-e), one can notice the following points; firstly, the appearance of second arc becomes clearer at higher temperatures and the frequency value at which the second arc appears increases with increasing temperature; second, the center of arc lying below the horizontal axis which suggests the non-Debye relaxation; third, the $G$ resistance decreases with increasing temperature. The $G$ resistance can be estimated from the intercept of first arc at the horizontal axis[34]. These results indicate that grain impedance lies in M$\Omega$ range except at 300K and at higher frequency, the grain contribution dominates. Further, in case of leaky (semiconducting) sample, due to higher frequency limit, it is difficult to see the $G$ contributions which will appear only at higher frequencies and one can assume the presence of single arc (as looks in the un-zoomed view of Nyquist plot for 270K and 300K in limited frequency window) which is associated to $G_b$ contributions rather than $G$. In such situations one has to see the intercept of Nyquist curve on $Z'$ axis[37]. Figure 4e demonstrates the Nyquist curves at higher frequencies for 300K, which clearly shows the appearance of second arc around ~68kHz. These observations suggest that at 300K, the dielectric behavior at low frequencies is mainly due to $G_b$ contributions whereas at higher frequencies (>68kHz), it is due to the $G$ contributions. However, in the absence of high frequency data, the amplitude of signal voltage (ac or dc) is also very useful tool to see the effect of grain boundaries. At low signal amplitudes, the $G_b$ impedance dominates the signal amplitude whereas at higher input amplitudes, the grain boundaries become less important because of its non-linear response[34,38]. We have, therefore, measured $Z'$ vs $f$ behavior at different applied ac amplitudes for different temperatures to further distinguish between the $G$ and $G_b$ contributions. From Fig. 4f, one can see the effect of different applied ac amplitude voltages on grain boundaries. Similar measurements were also performed at low temperatures ($T \leq 160K$) but we did not notice any divergence as at these temperatures, the sample is highly insulating and the overall contribution to the dielectric behaviour is coming out from grain only. Thus from Fig.4, one can clearly distinguished the frequency range at which the grain or grain boundary response dominates.

In impedance spectra, high resistive part is mostly dominating, whereas in the case of electric modulus one can see the contributions from different capacitors even if their resistances are comparable. Hence it's quite possible that sometime one can observe only a single arc in impedance plots whereas two peaks can be observed in the corresponding modulus spectra or vice-versa[39]. Therefore, one has to analyze the data in different ways, as it reveals which component (resistance or capacitance) is dominating in the observed dielectric behavior. Electric modulus is defined as[34]

$$M^* = \frac{1}{\varepsilon^*} \; ; \; M^* = M' + M'' = i\omega C_0 Z^* \qquad (3)$$

Real and imaginary part of electric modulus ($M^*$) can be written as

$$M' = \frac{\varepsilon'}{\varepsilon'^2 + \varepsilon''^2} \; ; \; M'' = \frac{\varepsilon''}{\varepsilon'^2 + \varepsilon''^2} \qquad (4)$$

where $M'$ and $M''$ are the real and imaginary parts of the complex electric modulus.

The $M'$ and $M''$ are plotted as a function of frequency at different temperatures in Fig. 5(a-b). The value of $M'$ is very small in the low frequency region, but increases with increasing frequency at all temperatures. The $M''$ vs $f$ curves (see Fig. 5b) shows a broad peak which shifts towards higher frequency with increasing temperature. The maximum angular frequency $\omega_{max}$ (=$2\pi f_{max}$) corresponds to $M''_{max}$ (maximum value of $M''$) is the reciprocal of the relaxation time, $\tau_M$[34] as:

$$\omega_{max.} = \frac{1}{\tau_M} \tag{5}$$

where $\tau_M$ is the relaxation time calculated from electric modulus. At different temperature this $\omega_{max}$ is different and increases with increasing temperature (see Fig. 5b). We have analyzed this data using equation (1) as done from Fig. 2b. Here the difference is that we are taking $\omega_{max}$ from isothermal measurements rather than temperature dependence at constant frequency. The fitted value of $\tau_o$ and $E_a$ (from modulus spectra) using equation (1) are nearly $\approx 1(\pm 0.7) \times 10^{-12}$s and $\approx 119$meV($\pm 3$), respectively, which are in well agreement with the value obtained from temperature dependent dielectric results (Fig. 2). We have also plotted the $M''/M''_{max}$ against $f/f_{max}$ (see Fig. 5c) for different temperatures. All these curves overlap in one master curve over most of the frequency region but slightly deviate at higher frequency side for few temperatures. This suggests that mostly single type of relaxation mechanism is following through all temperature regimes but at higher frequencies, it becomes temperature dependent to some extent. Further the full width half maxima value is 1.2 decades which is greater than 1.14 decades for Debye relaxation and proves that this relaxation follows non-Debye type behavior[34]. Further the master curve shows slightly asymmetric behavior at right hand side of the peak which again indicates that the relaxation dynamics is slightly temperature dependent at higher frequency side. Such asymmetric behavior is mostly governed by chemical disorder in the materials[34,38]. In our case this disorder may be due to the random distribution of Fe and Ti. Similar situation is also observed in other disorder systems.

The frequency dependent ac conductivity ($\sigma_{ac}$) is calculated from permittivity and $\tan\delta$ by using the formula[34]

$$\sigma_{ac}(\omega) = \varepsilon_o \omega \varepsilon' \tan\delta \tag{6}$$

Here $\varepsilon_o$ is the permittivity of free space (~$8.854 \times 10^{-12}$ F/m). Figure 6a shows the variation of $\sigma_{ac}$ with frequency at different temperature. From Fig. 6a, it is clear that, for $T>230$K, the linear region of $\sigma_{ac}$ i.e. the second universal (SU)[34,38] behavior of dielectric constant of FTO lies outside the measured frequency window.

The ac conductivity of any dielectric or semiconducting material can be expressed in term of frequency[34] as,

$$\sigma_{ac}(\omega) = A\omega^s \tag{7}$$

Here $A$ is the pre-exponential factor and $s$ is the frequency exponent whose value lies between 0 and 1. This formula is known as Jonscher's power law.

We have extracted the value of exponent '$s$' using ac conductivity, as shown in Fig. 6a, using the well established method[32,40,41]. This method implies the fitting of ac conductivity data using equation (7). We have fitted only the high frequency region (nearly linear) of ac conductivity where the '$s$' value remains frequency independent[42]. It is also clear from Fig. 6(a) that the linear region of ac conductivity shifts towards higher frequency with increasing temperature. We are unable to extract the '$s$' value for 270K and 300K because the linear region lies outside the measured frequency window. Here the T dependence of '$s$' can be expressed as[41]

$$s = 1 - \frac{6K_BT}{W_m} \tag{8}$$

where $W_m$ is the maximum barrier height. The variation of '$s$' with $T$ gives an idea about the conduction mechanism. For example, for quantum mechanical tunneling (QMT) '$s$' will be $T$ independent, for small polaron, '$s$' will increase with increasing $T$ whereas for CBH, '$s$' should increase with decreasing $T$ and approaches unity as $T$ tends to 0K in contrast to other hoping model[41-43]. The estimated variation of exponent '$s$' as a function of $T$ is shown in Fig. 6b. From Fig. 6b, it is clear that for FTO, '$s$' decreases with increasing $T$ which suggests that conduction mechanism supports CBH model[41-44]. The value of $W_m$ has been calculated by fitting the '$s$' versus $T$ curve with equation(8). The estimated value is 164(±2)meV which is of the same order as the activation energy value calculated from temperature dependent permittivity.

## IV. CONCLUSION

In summary, our comprehensive studies on FTO dielectric behavior illustrate the presence of non-Debye type dielectric relaxation around (220-270K), which is mainly associated with polaron hopping. The complex dielectric, impedance, electric modulus and ac conductivity have been analyzed to understand the $G$ and $G_b$ contributions. Our analysis illustrates that at low temperatures in the measured frequency range, the observed dielectric behavior is associated from $G$ contributions only, whereas the $G_b$ contributions start appearing around 175K (at low frequency). The observed scaling of isothermal electric modulus at different temperatures indicates that the transport dynamics is almost temperature independent across the relaxation regime. The ac conductivity analysis shows that the conduction mechanism is mainly governed by thermally activated process through polaron hopping and follows correlated barrier hopping model.

**Figure captions:**

**FIG. 1.** The schematic arrangement of $FeO_6$ and $TiO_6$ octahedra in the crystal structure of $Fe_2TiO_5$.

**FIG. 2.** Temperature dependence of (a) $\varepsilon'$ and (b) $\tan\delta$ at different frequencies during warming; inset in (b) shows the Arrhenius plot ($\tau$ versus temperature).

**FIG. 3.** Isothermal frequency variation of (a) $\varepsilon'$, (b) $\tan\delta$, (c) $Z'$ and (d) $Z''$ at different temperatures.

**FIG. 4.** Nyquist plot at (a) 160K, (b) 175K, (c) 220K, (d) 270K, (e) 300K at 1V signal amplitude and (f) 300K at different signal amplitudes and these values are mentioned in the figure. The arrows indicate the direction of increasing frequency.

**FIG. 5.** Frequency dependence of (a) $M'$ and (b) $M''$ at different temperatures. (c) Normalized curves of electric modulus at different temperatures; the inset shows the Arrhenius plot of modulus spectra.

**FIG. 6.** (a) Isothermal frequency dependence of real part of ac conductivity at different temperatures. (b) Variation of frequency exponent '$s$' with temperature including error bars.

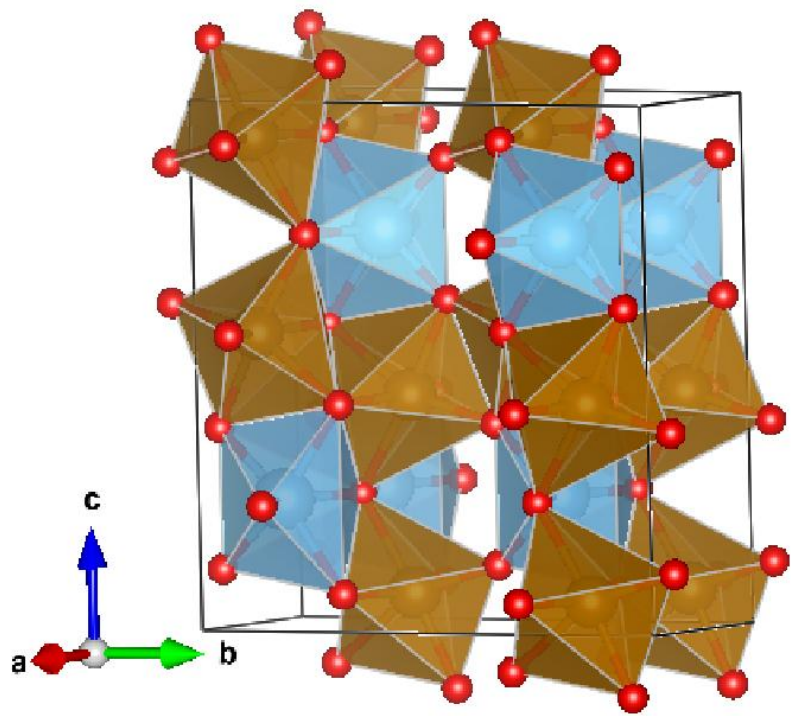

**Figure 1**

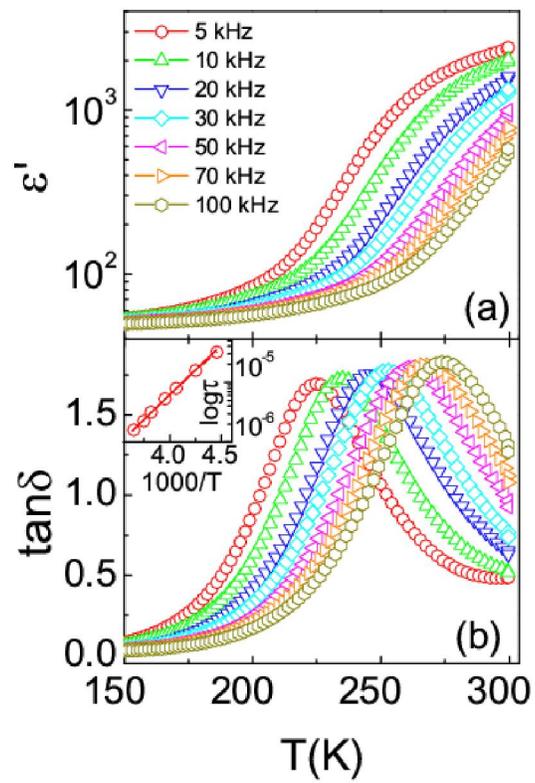

**Figure 2**

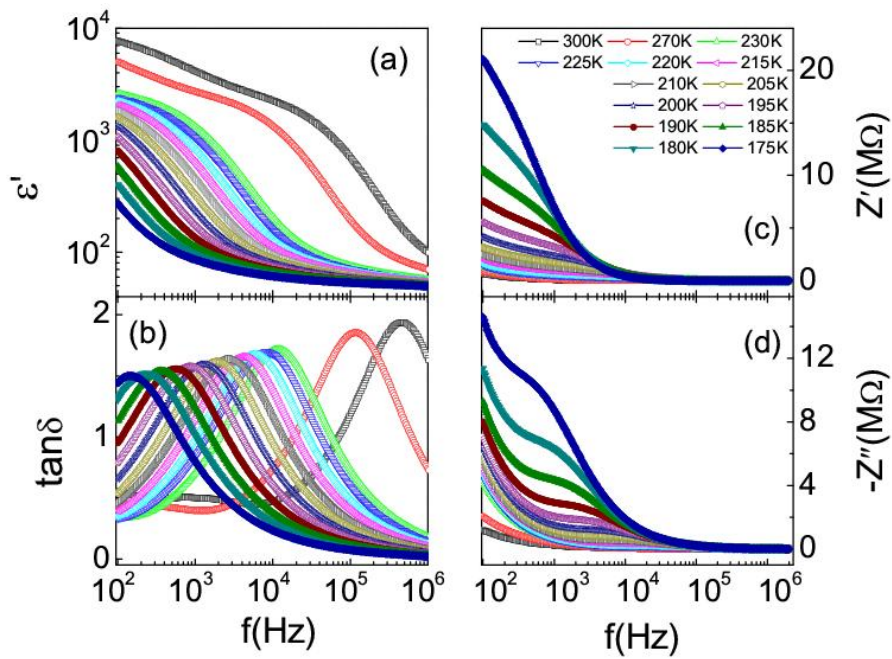

**Figure 3**

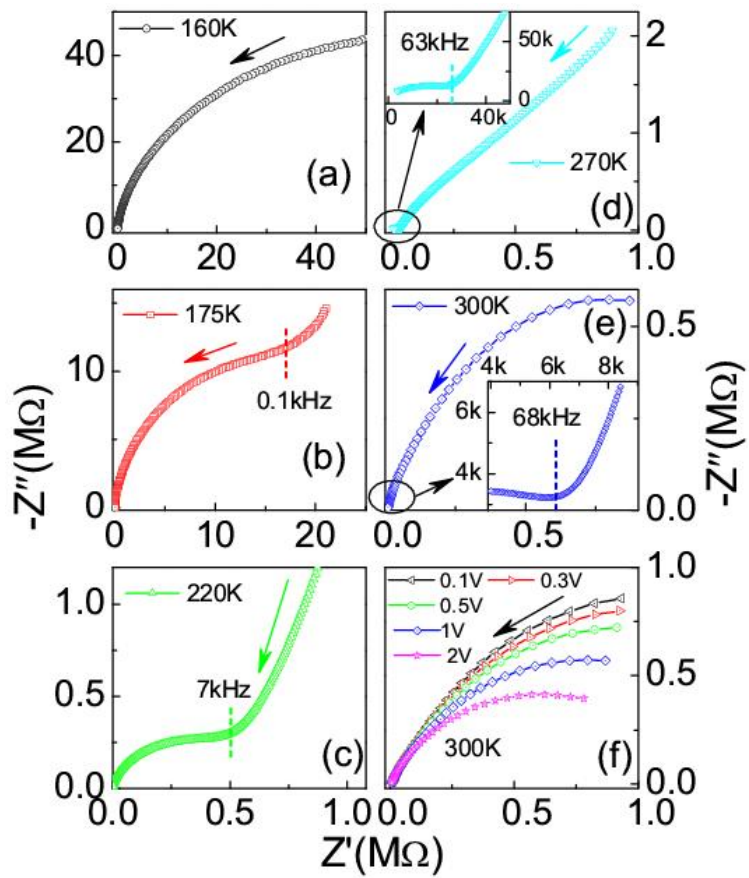

**Figure 4**

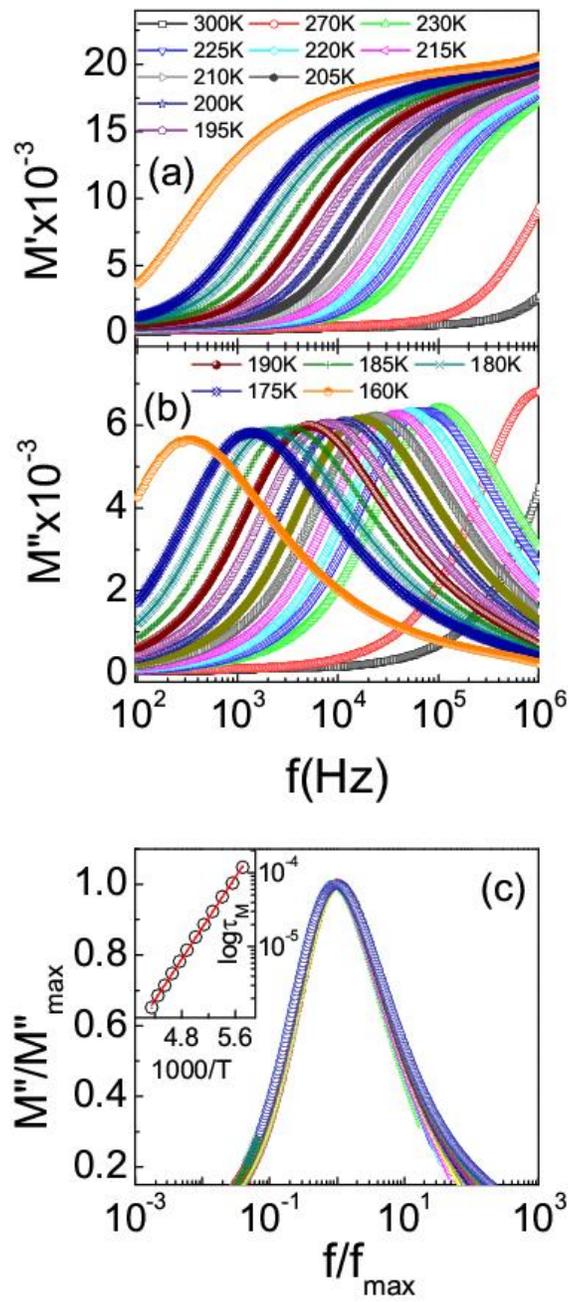

**Figure 5**

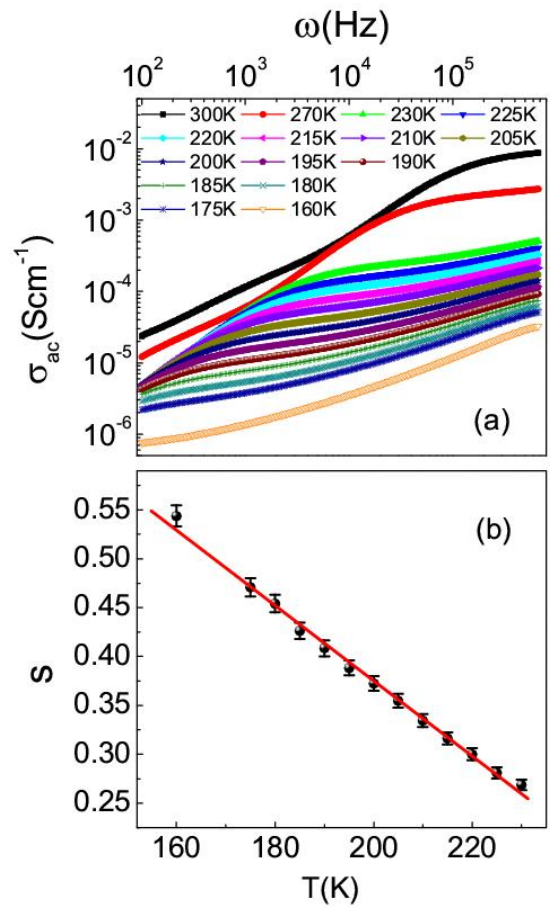

**Figure 6**